\documentclass[review]{elsarticle}

\usepackage{epsfig}

\usepackage{amsfonts}
\usepackage{mathrsfs}
\usepackage{bm}

\biboptions{sort&compress}

\usepackage[usenames,dvipsnames]{xcolor}

\makeatletter
\def\ps@pprintTitle{%
  \let\@oddhead\@empty
  \let\@evenhead\@empty
  \def\@oddfoot{\reset@font\hfil\thepage\hfil}
  \let\@evenfoot\@oddfoot
}
\makeatother

\def\be{\begin{equation}}
\def\ee{\end{equation}}
\def\bea{\begin{eqnarray}}
\def\eea{\end{eqnarray}}
\def\bfl{\begin{flushleft}}
\def\efl{\end{flushleft}}
\def\bfr{\begin{flushright}}
\def\efr{\end{flushright}}
\def\bc{\begin{center}}
\def\ec{\end{center}}
\def\ben{\begin{enumerate}}
\def\een{\end{enumerate}}
\def\bit{\begin{itemize}}
\def\eit{\end{itemize}}

\def\dzn{,\kern-0.1em,}

\journal{Physica B: Condensed Matter}

\begin{document}
\begin{frontmatter}

 %\title{Neel temperature\tnoteref{label1}}

\title{Effects of frustration and cyclic exchange on the spin-$1/2$ Heisenberg antiferromagnet within the self-consistent spin-wave theory}

\author{Milica S. Rutonjski \corref{cor}}
\ead{milica.rutonjski@df.uns.ac.rs}

\author{Maja B. Berovi\' c, Milica V. Pavkov-Hrvojevi\' c}

\cortext[cor]{Corresponding author}

\address{Department of Physics, Faculty of Sciences, University of Novi Sad,
Trg Dositeja Obradovi\' ca 4, Novi Sad, Serbia}

\begin{abstract}
The relevance of the quasi-two-dimensional spin-$1/2$ frustrated quantum antiferromagnet due to its possibility of modelling the high-temperature superconducting parent compounds has resulted in numerous theoretical and experimental studies. This paper presents a detailed research of the influence of the varying exchange interactions on the model magnetic properties within the framework of self-consistent spin-wave theory based on Dyson-Maleev representation. Beside the nearest neighbour interaction within the plane, the planar frustration up to the third nearest neighbours, cyclic interaction and the interlayer coupling are taken into account. The detailed description of the elementary spin excitations, staggered magnetization, spin-wave velocity renormalization factor and ground-state energy is given. The results are compared to the predictions of the linear spin-wave theory and when possible also to the second-order perturbative spin-wave expansion results. Finally, having at our disposal improved experimental results for the in-plane spin-wave dispersion in high-$T_c$ copper oxide $\mbox{La}_2\mbox{CuO}_4$, the self-consistent spin-wave theory is applied to that compound in order to correct earlier obtained set of exchange parameters and high temperature spin-wave dispersion.

\end{abstract}

%-----------------------------------------------------------------------------%


\end{frontmatter}

\section{Introduction}

\par The importance of the quasi-two-dimensional (2D) antiferromagnetically (AFM) ordered systems, corroborated by the fact that the high-temperature superconducting cuprate parent compounds, $\mbox{La}_2\mbox{CuO}_4$ for instance, belong to that class of systems, highly justifies their decades-long presence in the scientific literature. The interest is especially devoted to their magnetic properties in order to elucidate the underlying microscopic mechanism of the high-temperature superconductivity. Though the 2D spin-$1/2$ Heisenberg model with only the nearest neighbour (NN) interaction coupling $J$ was frequently used as the research initial point \cite{manusakis}, numerous studies, both theoretical and experimental, have in the meanwhile indicated the significance of taking into account other exchange interactions (see \cite{staricoldea,novicoldea,starikatanin,majumdarIoPscsw,majumdar_bez_cikl,majumdar_sa_cikl,majumdarIoPspinwaves,majumdarIoPmagnetization} and references therein), leading to the study of the frustrated spin systems \cite{diep}. Most emphasized is the relevance of the planar frustration, induced by the next-(NNN) or next-next-nearest (NNNN) neighbour interaction $J_{2/3}$, as well as the cyclic exchange $J_c$ comprising four-spin plaquette interaction, often recognized as the second strongest exchange interaction in the system. 
\par The self-consistent spin-wave theory (SCSWT) with its numerous versions is widely used for the analysis of the Heisenberg layered magnets (see \cite{zitomirski,stariirkin,irkin,kar} and references therein). Our research was primarily motivated by the study of the three-dimensional (3D) antiferromagnet with the planar frustration and cyclic exchange interaction within the SCSWT performed in Ref. \cite{starikatanin}. The theory was applied strictly to $\mbox{La}_2\mbox{CuO}_4$, without giving a detailed study of the dependence of the model magnetic properties on the varying values of the exchange interactions contained in Hamiltonian. This, however, may be interesting to investigate, especially since it is experimentally possible to vary the frustration ratio $J_2/J$ by applying high pressures on the system \cite{pavarini}. The present paper is intended to be a comprehensive study on the influence of planar frustration ratio, cyclic exchange and interplanar exchange interaction on the spin-wave spectrum, ground-state sublattice magnetization, spin-wave velocity renormalization factor and ground-state energy of the three-dimensional (3D) quantum antiferromagnet with $S=1/2$ within the self-consistent spin-wave theory. We shall determine the exchange parameter region where the ground state of the N\' eel type exists. The existence of the other long-range ordering patterns (columnar phase, for instance), as well as the intermediate quantum disordered phases, will not be the issue of this paper. The SCSW approach will be compared to the linear spin-wave theory (LSWT) results, as well as the results of second-order perturbative spin-wave expansion from \cite{majumdar_bez_cikl,majumdar_sa_cikl,majumdarIoPspinwaves,majumdarIoPmagnetization}. The influence of the third nearest neighbour interaction on sublattice magnetization will be analyzed, which to our knowledge has not been earlier discussed within the framework of SCSWT applied on the above introduced model. Special attention will be devoted to the high-$T_c$ superconducting parent compound $\mbox{La}_2\mbox{CuO}_4$, thoroughly studied in the literature. Using the improved experimental in-plane spin-wave dispersion \cite{novicoldea} we shall find the corrected exchange parameter set, whereby we shall calculate spin-wave dispersion at high temperatures and compare it both to the experiment from the Ref. \cite{staricoldea}, as well as to the numerically obtained spectrum from \cite{starikatanin}.
\par The paper is organized as follows: in Section 2 we present the model Hamiltonian and introduce the dominant exchange interactions. In Section 3 the Hamiltonian is mapped into the bosonic form based on the Dyson-Maleev (DM) transformation. Mean-field decoupling of the fourth-order terms is performed and main expressions for the quantities to be analyzed are derived by making use of the self-consistent spin-wave theory. The numerically calculated results are plotted and analyzed in detail in Sec. 4. The conclusions are shortly stated in Sec. 5. 

\section{Model Hamiltonian}

\par The dominant interactions for the $S=1/2$ antiferromagnet on the tetragonal lattice are presented in Figure 1 
and comprised in the following Heisenberg Hamiltonian:
$$ \hat{H}=J\sum_{\vec{n}_a,\vec{\delta}_1}\hat{\vec{S}}_{\vec{n}_a}\cdot \hat{\vec{S}}_{\vec{n}_a+\vec{\delta}_1}+\frac{1}{2}\sum_{\vec{n}_{\alpha}\atop(\alpha=a,b)}\sum_{\vec{\delta}_i\atop (i=2,3)}J_i\,\hat{\vec{S}}_{\vec{n}_{\alpha}}\cdot \hat{\vec{S}}_{\vec{n}_{\alpha}+\vec{\delta}_i}+$$
$$ +J_c \sum_{\vec{n}_{\alpha}\atop(\alpha=a,b)} \left[ \left(\hat{\vec{S}}_{\vec{n}_{\alpha}}\cdot \hat{\vec{S}}_{\vec{n}_{\alpha}+\vec{d}_1}\right)\left( \hat{\vec{S}}_{\vec{n}_{\alpha}+\vec{d}_2}\cdot \hat{\vec{S}}_{\vec{n}_{\alpha}+\vec{d}_3}\right)+\left(\hat{\vec{S}}_{\vec{n}_{\alpha}}\cdot \hat{\vec{S}}_{\vec{n}_{\alpha}+\vec{d}_3}\right)\left( \hat{\vec{S}}_{\vec{n}_{\alpha}+\vec{d}_1}\cdot \hat{\vec{S}}_{\vec{n}_{\alpha}+\vec{d}_2}\right)-\right.$$ \be\left.-\left(\hat{\vec{S}}_{\vec{n}_{\alpha}}\cdot \hat{\vec{S}}_{\vec{n}_{\alpha}+\vec{d}_2}\right)\left( \hat{\vec{S}}_{\vec{n}_{\alpha}+\vec{d}_1}\cdot \hat{\vec{S}}_{\vec{n}_{\alpha}+\vec{d}_3}\right)\right]+ 
J_{\bot}\sum_{\vec{n}_a,\vec{\delta}_{\bot}^{ab}}\hat{\vec{S}}_{\vec{n}_a}\cdot \hat{\vec{S}}_{\vec{n}_a+\vec{\delta}_{\bot}^{ab}}+\frac{J_{\bot}}{2}\sum_{\vec{n}_{\alpha},\delta_{\bot}^{\alpha\alpha}}\hat{\vec{S}}_{\vec{n}_{\alpha}}\cdot \hat{\vec{S}}_{\vec{n}_{\alpha}+\vec{\delta}_{\bot}^{\alpha\alpha}}\,.\label{hamiltonian1}\ee
The first two terms include the interactions between the first, second and third nearest neighbours within the plane. The position of the spin in the sublattice $\alpha$ ($\alpha=a,b$) is denoted by $\vec{n}_{\alpha}$, while the vectors connecting the given spin and its corresponding neighbours are denoted by $\vec{\delta}_i$ ($i=1,2,3$).  The third term presents the cyclic exchange interaction, described by the exchange integral $J_c$. The vectors connecting the spins in a plaquette read: $\vec{d}_1=a\,(\vec{e}_x+\vec{e}_y)$, $\vec{d}_2=a\,\vec{e}_x$ and $\vec{d}_3=a\,(\vec{e}_x-\vec{e}_y)$. The interplanar interaction is described by the last two terms, where the vectors $\vec{\delta}_{\bot}^{ab/\alpha\alpha}$ designate the vectors connecting the antiferro-/ferromagnetically coupled nearest neighbour spins in adjacent planes. All interactions are assumed to be antiferromagnetic, i.e. $J$, $J_2$, $J_3$, $J_c$, $J_{\bot}$ $>0$. Contrary to Refs. \cite{majumdarIoPscsw,majumdar_bez_cikl,majumdar_sa_cikl,majumdarIoPspinwaves,majumdarIoPmagnetization}, we observe that Hamiltonian (\ref{hamiltonian1}) includes the next-next-nearest neighbour planar interaction, enabling us to incorporate in the present study the influence of this interaction in the N\' eel phase.

\bc{\includegraphics{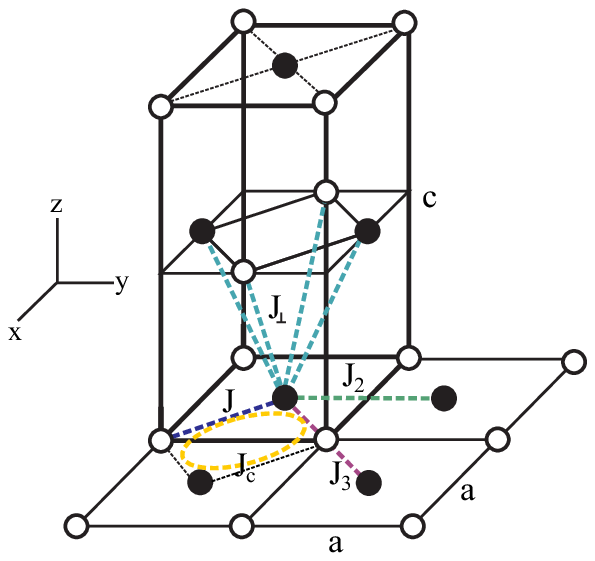} \\ {\small Figure 1: (Color online) Magnetic unit cell (bold solid line) of 3D antiferromagnet with exchange interactions labeled. Two different orientations of spins are denoted by open and solid circles.}}\ec

It is common to define the fundamental energy scale by the NN exchange interaction $J$. Therefore, we introduce the dimensionless ratios and study their influence on the spin-wave spectrum, ground-state sublattice magnetization etc. These quantities are the frustration ratio $\lambda_{2/3}=J_{2/3}/J$, the ratio $\lambda_c=J_c/J$ parametrizing the relative strength of the cyclic exchange and the ratio $\lambda_{\bot}=J_{\bot}/J$ parametrizing the relative strength of the interplanar interaction.

%------------------------------------------------------------------------------------------------------------------------------------------------------

\section{Spin-wave spectrum, magnetization and related quantities}

In order to determine the spin-wave spectrum, we apply a variant of the non-linear spin-wave theory \cite{irkinuspehi}. The spin operators in Hamiltonian (\ref{hamiltonian1}) are first written in the local coordinate system \cite{PRB1mi} and then expressed as boson operators using the well-known Dyson-Maleev transformation in the form
$$ \hspace*{-1.5cm}\hat{\sigma}^+_{\vec{n}_{\alpha}}=\sqrt{2S}\hat{a}_{\vec{n}_{\alpha}}\,,\,\,\,\,
\hat{\sigma}^-_{\vec{n}_{\alpha}}=\sqrt{2S}\hat{a}^{\dagger}_{\vec{n}_{\alpha}}-\frac{1}{\sqrt{2S}}\hat{a}^{\dagger}_{\vec{n}_{\alpha}}\hat{a}^{\dagger}_{\vec{n}_{\alpha}}\hat{a}_{\vec{n}_{\alpha}}
\,,$$
\be\hat{\sigma}^z=S-\hat{a}^{\dagger}_{\vec{n}_{\alpha}}\hat{a}_{\vec{n}_{\alpha}}\,,\hspace{0.4cm}\alpha=a,b\,.\label{dajson}\ee
In terms of these operators, the Hamiltonian (\ref{hamiltonian1}) is mapped into an equivalent Hamiltonian of interacting bosons, which may be written as
\be \hat{H}=H_0+\hat{H}_2+\hat{H}_4\,,\label{hamiltonian2}\ee
where $H_0$ presents the classical ground-state energy, $\hat{H}_2$ denotes the bilinear part of the Hamiltonian which corresponds to the non-interacting spin waves (LSW), while $\hat{H}_4$ describes the interaction among spin waves induced via transformations (\ref{dajson}) and consists of the terms quartic in Bose operators. The terms of the higher order are immediately neglected. We then simplify the Hamiltonian (\ref{hamiltonian1}) in the spirit of the Wick's theorem \cite{fetervaleka}, in the manner similar to the one implemented in Ref. \cite{stanekPRB}. The manner in which the quartic terms are evaluated is illustrated on the following example:
$$\hat{a}_{\vec{n}_a}\hat{b}^{\dagger}_{\vec{n}_a+\vec{\delta}_1}\hat{b}^{\dagger}_{\vec{n}_a+\vec{\delta}_1}\hat{b}_{\vec{n}_a+\vec{\delta}_1}\approx  2\langle\hat{a}_{\vec{n}_a}\hat{b}^{\dagger}_{\vec{n}_a+\vec{\delta}_1}\rangle\,\hat{b}^{\dagger}_{\vec{n}_a+\vec{\delta}_1}\hat{b}_{\vec{n}_a+\vec{\delta}_1}+\langle\hat{a}_{\vec{n}_a}\hat{b}_{\vec{n}_a+\vec{\delta}_1}\rangle\,\hat{b}^{\dagger}_{\vec{n}_a+\vec{\delta}_1}\hat{b}^{\dagger}_{\vec{n}_a+\vec{\delta}_1}+$$
\be +2\langle\hat{b}^{\dagger}_{\vec{n}_a+\vec{\delta}_1}\hat{b}_{\vec{n}_a+\vec{\delta}_1}\rangle\,\hat{a}_{\vec{n}_a}\hat{b}^{\dagger}_{\vec{n}_a+\vec{\delta}_1}+\langle\hat{b}^{\dagger}_{\vec{n}_a+\vec{\delta}_1}\hat{b}^{\dagger}_{\vec{n}_a+\vec{\delta}_1}\rangle\,\hat{a}_{\vec{n}_a}\hat{b}_{\vec{n}_a+\vec{\delta}_1}\,,\label{dekuplovanje}\ee
where the single brackets $\langle ...\rangle$ indicate averages over the canonical ensemble at temperature $T$. As a consequence of DM transformation, the obtained Hamiltonian is non-Hermitian. Therefore, we symmetrize it by adding the Hermitian conjugate terms to the non-Hermitian ones \cite{flax}. After decoupling procedure is completed, we perform the Fourier transform and obtain the following bosonic Hamiltonian
\begin{eqnarray}\hat{H}&=&-JS^2 z \frac N2+J_2 S^2\frac N2 z+J_3 S^2\frac N2 z+J_c S^4N+\nonumber\\
& &+\sum_{\vec{k}}\epsilon(\vec{k})\left(\hat{a}^{\dagger}_{\vec{k}}\hat{a}_{\vec{k}}+\hat{b}^{\dagger}_{\vec{k}}\hat{b}_{\vec{k}}\right)+\sum_{\vec{k}}\alpha(\vec{k})\left(\hat{a}^{\dagger}_{\vec{k}}\hat{b}^{\dagger}_{\vec{-k}}+\hat{a}_{\vec{k}}\hat{b}_{\vec{-k}}\right)\,.\label{hamiltonian3}\end{eqnarray} 
Here, $z$ denotes the coordination number and equals four in all cases, $N$ presents the number of lattice sites, while the quantities $\epsilon(\vec{k})$ and $\alpha(\vec{k})$ are given by the expressions
\begin{eqnarray}\epsilon(\vec{k})&=&JSz\left[ \Gamma_1-\lambda_2 \Gamma_2 (1-\gamma_2(\vec{k}))-\lambda_2\Gamma_3 (1-\gamma_3(\vec{k}))-\right.\nonumber\\
& &-\left.\frac{\lambda_c}{4}(\Gamma_{c(1)}+\Gamma_{c(2)}\gamma_2(\vec{k}))+\lambda_{\bot}(\Gamma_{\bot}^{ab}-\Gamma_{\bot}^{aa}(1-\gamma_{\bot}^{aa}(\vec{k})))\right]\,,\label{epsilon}\end{eqnarray}
\be\alpha(\vec{k})=JSz\left[\Gamma_1\gamma_1(\vec{k})-\frac{\lambda_c}{2}\Gamma_{c(3)}\gamma_1(\vec{k})+\lambda_{\bot}\Gamma_{\bot}^{ab}\gamma_{\bot}^{ab}(\vec{k})\right]\,.\label{alfa}\ee
The quantities $\gamma(\vec{k})$ denote the structure factors with lattice constants set to unity:
$$ \gamma_1(\vec{k})=\mbox{cos}\frac{k_x}{2}\mbox{cos}\frac{k_y}{2}
\hspace{0.7cm}\gamma_2(\vec{k})=\frac12(\mbox{cos}\,k_x+\mbox{cos}\,k_y)\hspace{0.7cm} \gamma_3(\vec{k})=\mbox{cos}\,k_x\mbox{cos}\,k_y\,,$$
\be\gamma_{\bot}^{ab}=\mbox{cos}\frac{k_x}{2}\mbox{cos}\frac{k_z}{2}\hspace{0.7cm}\gamma_{\bot}^{aa}=\mbox{cos}\frac{k_y}{2}\mbox{cos}\frac{k_z}{2}\,,\label{game_geometrijske}\ee
while the quantities $\Gamma$ present the renormalization factors given by 
\begin{eqnarray}\Gamma_1 &=& 1-\frac1S\left[A(T)+D(T)\right]\nonumber\\
\Gamma_2&=&1-\frac1S\left[A(T)-L(T)\right]\nonumber\\ \Gamma_3&=&1-\frac1S\left[A(T)-M(T)\right]\nonumber\\
\Gamma_{c(1)}&=&1-\frac1S\left[3A(T)+3G(T)+6D(T)\right]\nonumber\\
\Gamma_{c(2)}&=&1-\frac1S\left[3A(T)+G(T)+4D(T)\right]\nonumber\\
\Gamma_{c(3)}&=&1-\frac1S\left[3A(T)+2G(T)+5D(T)\right]\nonumber\\
\Gamma_{\bot}^{ab}&=&1-\frac1S\left[A(T)+P(T)\right]\nonumber\\ 
\Gamma_{\bot}^{aa}&=&1-\frac1S\left[A(T)-Q(T)\right]\,.\label{renorm_faktori}\end{eqnarray}
In Bloch's approximation all these renormalization factors equal unity, while here due to the quantum and thermal fluctuations they are defined by the aforementioned expressions. The temperature functions which enter Eqs. (\ref{renorm_faktori}) are the following correlation functions
\begin{eqnarray} A(T)&=&\frac2N\sum_{\vec{k}}\langle\hat{a}^{\dagger}_{\vec{k}}\hat{a}_{\vec{k}}\rangle=\frac1N\sum_{\vec{k}}\left(\frac{\epsilon(\vec{k})}{E(\vec{k})}\mbox{coth}\frac{E(\vec{k})}{2\theta}-1\right)\nonumber\\
G(T)&=&\frac2N\sum_{\vec{k}}\langle\hat{a}^{\dagger}_{\vec{k}}\hat{a}_{\vec{k}}\rangle\mbox{cos}\,k_x=\frac1N\sum_{\vec{k}}\left(\frac{\epsilon(\vec{k})}{E(\vec{k})}\mbox{coth}\frac{E(\vec{k})}{2\theta}-1\right)\mbox{cos}\,k_x\nonumber\\
L(T)&=&\frac2N\sum_{\vec{k}}\langle\hat{a}^{\dagger}_{\vec{k}}\hat{a}_{\vec{k}}\rangle\gamma_2(\vec{k})=\frac1N\sum_{\vec{k}}\left(\frac{\epsilon(\vec{k})}{E(\vec{k})}\mbox{coth}\frac{E(\vec{k})}{2\theta}-1\right)\gamma_2(\vec{k})\nonumber\\
M(T)&=&\frac2N\sum_{\vec{k}}\langle\hat{a}^{\dagger}_{\vec{k}}\hat{a}_{\vec{k}}\rangle\gamma_3(\vec{k})=\frac1N\sum_{\vec{k}}\left(\frac{\epsilon(\vec{k})}{E(\vec{k})}\mbox{coth}\frac{E(\vec{k})}{2\theta}-1\right)\gamma_3(\vec{k})\nonumber\\
D(T)&=&\frac2N\sum_{\vec{k}}\langle\hat{a}_{\vec{k}}\hat{b}_{\vec{-k}}\rangle\gamma_1(\vec{k})\hspace{-0.1cm}=\hspace{-0.1cm}-\frac1N\sum_{\vec{k}}\sqrt{\frac{\epsilon^2(\vec{k})}{E^2(\vec{k})}-1}\,\gamma_1(\vec{k})\,\mbox{coth}\frac{E(\vec{k})}{2\theta}\nonumber\\
Q(T)&=&\frac2N\sum_{\vec{k}}\langle\hat{a}^{\dagger}_{\vec{k}}\hat{a}_{\vec{k}}\rangle\gamma_{\bot}^{aa}(\vec{k})=\frac1N\sum_{\vec{k}}\left(\frac{\epsilon(\vec{k})}{E(\vec{k})}\mbox{coth}\frac{E(\vec{k})}{2\theta}-1\right)\gamma_{\bot}^{aa}(\vec{k})\nonumber\\
P(T)&=&\frac2N\sum_{\vec{k}}\langle\hat{a}_{\vec{k}}\hat{b}_{\vec{-k}}\rangle\gamma_{\bot}^{ab}(\vec{k})\hspace{-0.1cm}=\hspace{-0.1cm}-\frac1N\sum_{\vec{k}}\sqrt{\frac{\epsilon^2(\vec{k})}{E^2(\vec{k})}-1}\,\gamma_{\bot}^{ab}(\vec{k})\,\mbox{coth}\frac{E(\vec{k})}{2\theta},\label{korelac_funkcije}\end{eqnarray}
where $\theta=k_B T$. The diagonalization of (\ref{hamiltonian3}) by making use of the standard Bogoliubov's transformation yields 
the spin-wave spectrum
\be E(\vec{k})=\sqrt{\epsilon^2(\vec{k})-\alpha^2(\vec{k})}\,,\label{spektar}\ee
while the ground-state energy per lattice site, with the quadratic and quartic corrections included, reads
\be E_0/N=-\frac 12 JS^2 z+\frac 12 J_2 S^2 z+\frac 12 J_3 S^2 z+J_c S^4+\frac 1N\sum_{\vec{k}}\left(E(\vec{k})-\epsilon(\vec{k})\right).\label{energ_osn_stanja}\ee The spectrum given by (\ref{spektar}) possesses the Goldstone mode, as expected due to the spin isotropy of the model. 
\par The sublattice magnetization in the absence of the external magnetic field is given by 
\be\langle\hat{S}^{z(a)}\rangle=S-\frac{1}{N_a}\sum_{\vec{k}}\left[\frac12 \frac{\epsilon(\vec{k})}{E(\vec{k})}\mbox{coth}\frac {E(\vec{k})}{2\theta}-\frac12\right]\,.
\label{mag_podresetke}\ee
\par Finally, we shall examine the spin-wave velocity renormalization factor $Z_c$ in 2D case. Namely, in the long-wave limit, the spin-wave dispersion vanishes linearly according to
\be E(\vec{k})\sim ck,\hspace{1cm}k=\sqrt{k_x^2+k_y^2}\rightarrow 0\,,\label{dugotalasna_granica}\ee
where $c$ presents the spin-wave velocity given by
\be c=2\sqrt{2}Z_c J S\label{brzina_spin_tal}\,.\ee
Therefrom, the renormalization factor $Z_c$ reads
\be Z_c=\sqrt{\left[\Gamma_1-\lambda_c S^2 (\Gamma_{c(1)}+\Gamma_{c(2)})\right]\left[\Gamma_1-2\lambda_2\Gamma_2-4\lambda_2\Gamma_3+2\lambda_c S^2 (\Gamma_{c(2)}-\Gamma_{c(3)})\right]}\,.\label{renorm_faktor}\ee

All the numerical calculations will be performed based on upper expressions. 
%--------------------------------------------------------------------------------------------------------------------------------------

\section{Analysis of results}

\subsection{Spin-wave energies}
\par In order to describe the ground-state behavior of the system we perform the calculation of the spin-wave spectrum for the 2D model within LSW and SCSW theory, based on numerical evaluation of Eqs. (\ref{epsilon}), (\ref{alfa}) and (\ref{spektar}), where we put $T=0$. In the LSW approach, $\Gamma$ quantities in last two expressions equal unity, while within the SCSW approach the cumbersome system of Eqs. (\ref{epsilon})-(\ref{korelac_funkcije}) has to be solved. Due to its self-consistency, the iterative procedure has to be applied. In Figure 2 we show the comparison between the spin-wave spectra along the high symmetry directions in the 2D AFM Brillouin zone (for the tetragonal phase) obtained within the LSW (blue line) and SCSW (red solid line) theory, for the relative frustration $\lambda_2=\lambda_3=0.1$ and cyclic exchange $\lambda_c=0$. 

\bc{\includegraphics{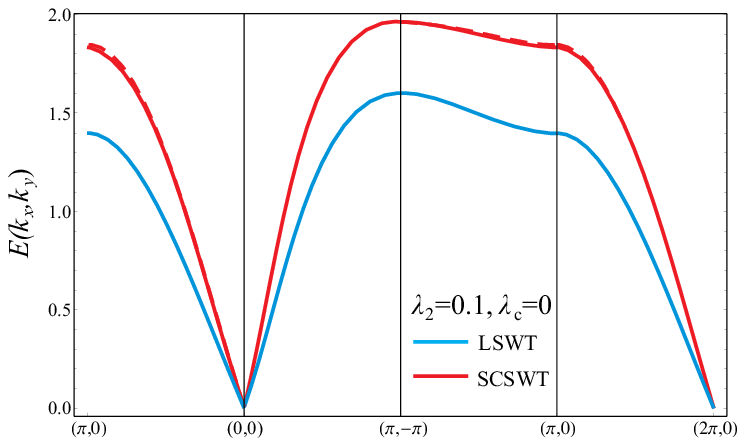} \\ {\small Figure 2: (Color online) Spin-wave energy $E(\vec{k})/J$ along the high symmetry directions in the 2D AFM Brillouin zone obtained from LSWT, SCSWT and first correction to LSWT (red dashed line), for the couplings $\lambda_2=0.1$ and $\lambda_c=0$.}}\ec

\noindent We also show the results obtained as the first correction to the LSWT result, i.e. by inserting the LSW results in the SCSW expressions, without performing the self-consistent procedure (presented in Fig. 1 with the red dashed line). If not stated otherwise, hereafter we assume that $\lambda_3$ equals $\lambda_2$ \cite{staricoldea,starikatanin} and do not emphasize the value of the former for brevity. While qualitatively LSWT and SCSWT results resemble each other, the SCSWT gives significant contribution to the entire dispersion curve. Further, it can be seen that the SCSWT results differ only slightly from the first correction to the LSWT and only in the vicinity of the Brillouin zone boundary. Other choices of exchange parameters confirm this observation, whereby this difference grows with the increase of $\lambda_c$, though very slowly, being less than $1\%$ for $\lambda_2=0.1$ and $\lambda_c=0.6$, which may bring us to the conclusion that the tedious procedure of solving the aforementioned system of equations iteratively is not justified. Though, the study of the ground-state sublattice magnetization and other quantities of interest will show that it is necessary to employ the SCSW theory in order to avoid the divergences which arise from taking the first correction to the LSWT only.
\par In Figure 3a) we present the LSW theory results for the spin-wave spectra for different values of parameters $\lambda_2$ and $\lambda_c$. 

\bc{\includegraphics[scale=0.75]{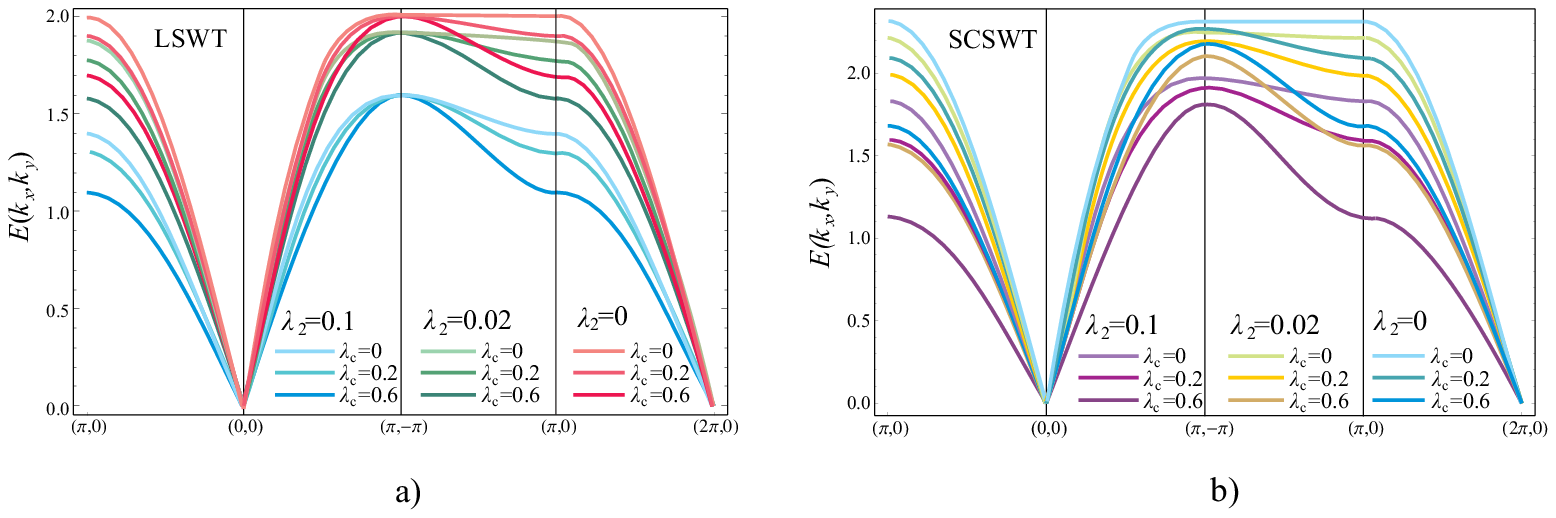} \\ {\small Figure 3: (Color online) a)Spin-wave energies $E(\vec{k})/J$ obtained within LSWT, for various values of parameters $\lambda_2$ and $\lambda_c$. b)Spin-wave energies $E(\vec{k})/J$ within SCSWT, for various values of parameters $\lambda_2$ and $\lambda_c$.}}\ec

\noindent It is obvious that with the increase of frustration, the magnon energies get lowered, with a clearly observable dip from $(\pi,-\pi)$ to $(\pi,0)$, meaning that the excitation energy at $(\pi,0)$ decreases more strongly than at $(\pi,-\pi)$. The dip grows with the frustration and is absent only for $\lambda_2=0$ and $\lambda_c=0$, as also stated in Ref. \cite{majumdar_sa_cikl}. The increase of cyclic exchange for fixed $\lambda_2$ supports the softening of $(\pi,0)$ mode, while the $(\pi,-\pi)$ mode remains unaffected, which presents the shortcoming of the LSW approach. This drawback is removed within the SCSWT, as seen in Figure 3b), where the evolution of the spin-wave spectra with varying parameters $\lambda_2$ and $\lambda_c$ is plotted. Within SCSWT the increase in $\lambda_c$ softens also the $(\pi,-\pi)$ mode, in accordance with \cite{majumdar_sa_cikl}. We observe that even in the absence of the cyclic interaction there exists a dip from $(\pi,-\pi)$ to $(\pi,0)$ (except for $\lambda_2=0$), which grows with frustration. Hence, in this case, SCSW theory gives result opposite to the one obtained within the second-order perturbative spin expansion \cite{majumdar_bez_cikl}, which predicts the dip of the same sign only for the substantial cyclic exchange, while for $\lambda_c=0$ the inverse dip from $(\pi,0)$ to $(\pi,-\pi)$ is obtained.

\subsection{Ground-state sublattice magnetization}

\par We proceed with the calculation of the ground-state sublattice magnetization, again for the 2D model at first, by making use of the expression (\ref{mag_podresetke}), whereby we again have to apply the iterative procedure. The LSWT and SCSWT results for the zero-temperature magnetization dependence on the frustration ratio are compared in Figure 4a). 

\bc{\includegraphics[scale=0.75]{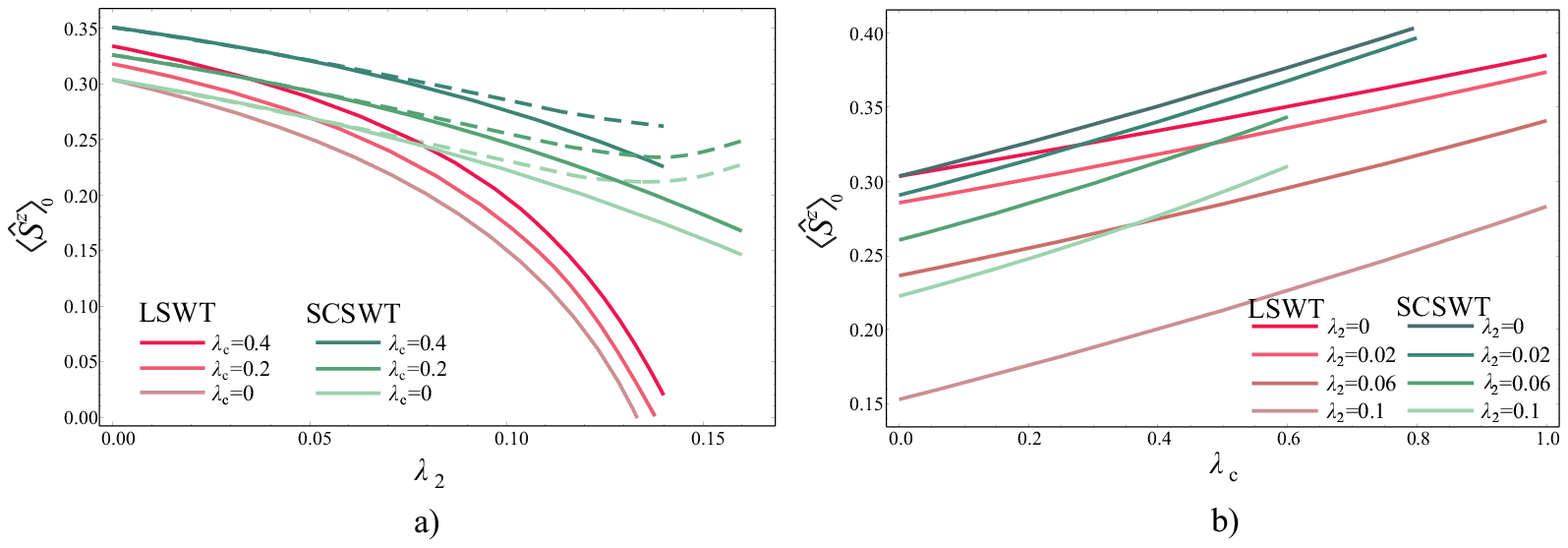} \\ {\small Figure 4: (Color online) a)The ground state sublattice magnetization dependence on $\lambda_2$ for different values of $\lambda_c$. The solid lines present the LSWT and SCSWT results. The dashed lines present the results of the first correction to the LSWT. b)The ground state sublattice magnetization vs. $\lambda_c$ for different values of $\lambda_2$ in LSW and SCSW approach.}}\ec

\noindent Since the presence of frustration disorders the system, the sublattice magnetization $\langle \hat{S}^z\rangle_0$ is gradually reduced by the increase of $\lambda_2$, i.e. the zero-point quantum fluctuations grow. The magnetization within SCSWT is enhanced and decreases more slowly compared to the LSWT predictions, in accordance with \cite{majumdarIoPscsw}. The apparent divergence which arises within the non-linear SW theory including the first correction to $1/S$ order only (shown in Figure 4a) by the dashed lines), is successfully removed by allowing for the SCSW approach. This conclusion agrees with those quoted in \cite{majumdar_bez_cikl,majumdar_sa_cikl}. The sublattice magnetization dependence on the cyclic exchange parameter is presented in Figure 4b). It can be seen that the cyclic exchange stabilizes the system, whereby the magnetization is enhanced within SCSWT. Besides, the difference between the LSW and SCSW theory results ($\langle \hat{S}^z\rangle_0^{\mbox{\tiny{LSWT}}}-\langle \hat{S}^z\rangle_0^{\mbox{\tiny{SCSWT}}}$) grows with the increase of the parameter $\lambda_2$.
\par It is important to emphasize that the domain of the parameters $\lambda_2$ and $\lambda_c$ where the ground state of the N\' eel type exists for the model defined by the Hamiltonian (\ref{hamiltonian1}) within SCSWT, is limited. The corresponding parameter region is shown in the phase diagram in Figure 5.
\bc{\includegraphics[scale=1]{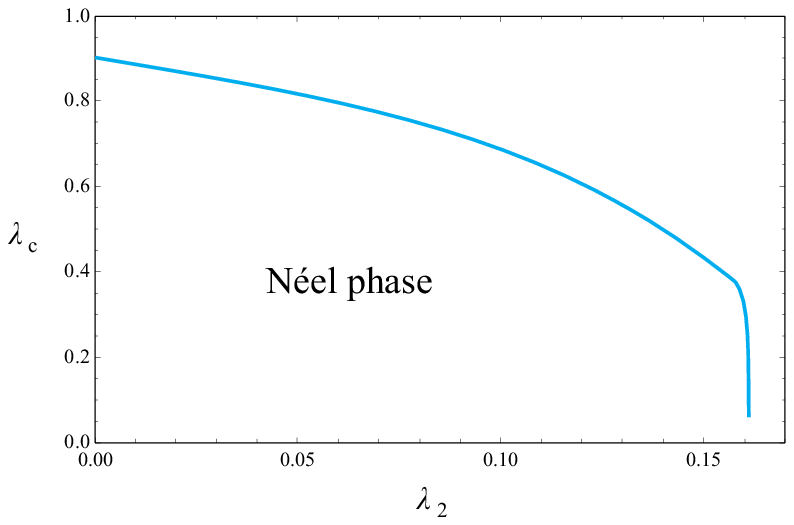} \\ {\small Figure 5: (Color online) The N\' eel phase parameter region in the $J$-$J_2$-$J_3$-$J_c$ model.}}\ec

\par Further, we study the influence of the third nearest neighbour interaction on the sublattice magnetization, in order to estimate its contribution to the ground-state sublattice magnetization. Though the NNNN interaction was included in the Hamiltonian in Ref. \cite{starikatanin}, its influence on the model magnetic properties was not separately discussed. The comparison between the LSWT and SCSWT results for the cases with and without the NNNN interaction (for simplicity in the absence of cyclic exchange) are presented in Figure 6a). 

\bc{\includegraphics[scale=0.7]{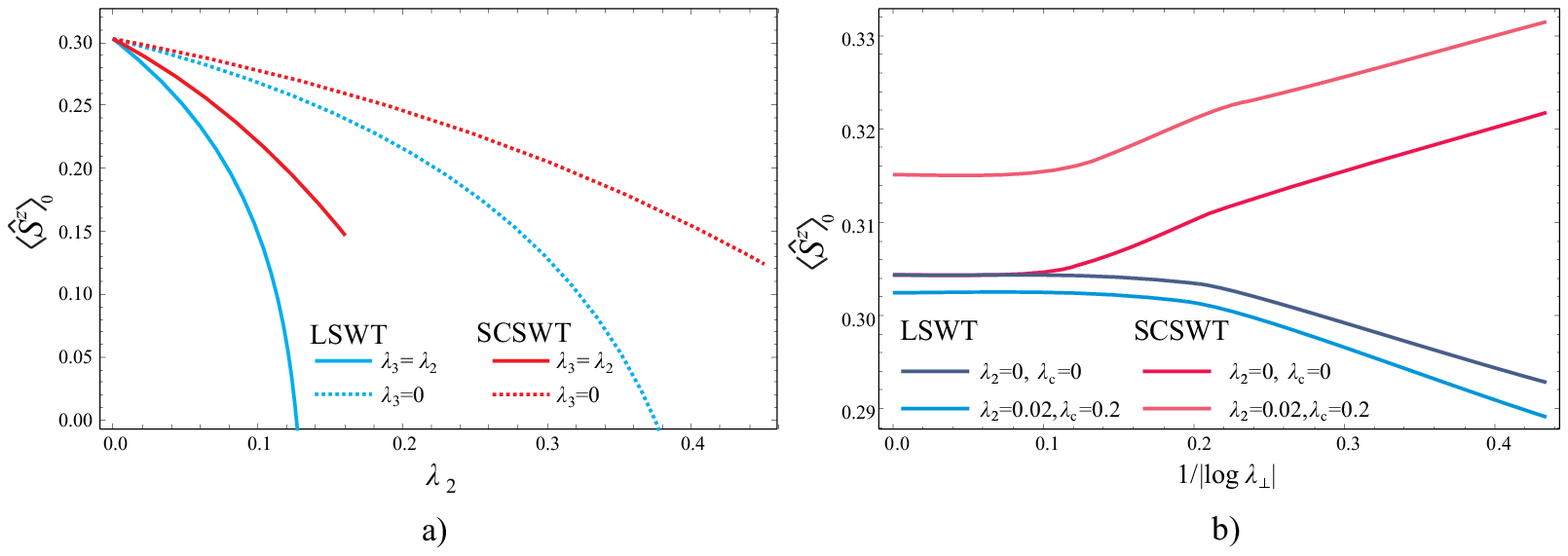} \\ {\small Figure 6: (Color online) a)The comparison of ground-state sublattice magnetization dependence upon $\lambda_2$ for $\lambda_3=0$ (dotted lines) and $\lambda_3=\lambda_2$ (solid lines). For simplicity, $\lambda_c=0$. b)The ground-state sublattice magnetization dependence on the parameter $\lambda_{\bot}$, for two choices of parameters $\lambda_2$ and $\lambda_c$.}}\ec

\noindent Evidently, with the presence of the non-vanishing parameter $\lambda_3$ the frustration in the antiferromagnet grows, additionally destabilizing the system. For instance, within the LSWT, the critical value of $\lambda_2$ at which the AFM order is destroyed, equals approximately $0.38$ in the absence of NNNN interaction, which presents the result also stated in \cite{majumdar_bez_cikl}, while it falls to a much lower value of $0.13$ for $\lambda_3\neq 0$. The SCSWT results also corroborate this observation.
\par Next, we allow for the interplanar interaction between the adjacent planes which complicates the analysis significantly, since it introduces the additional frustration into the system (see Figure 1). In order to examine the effects of the three-dimensionality on the behavior of the system, we determine the influence of the parameter $\lambda_{\bot}$ on the ground-state sublattice magnetization. The calculated results are plotted in Figure 6b) both for the simple model with $\lambda_2=\lambda_3=\lambda_c=0$ and the model with non-vanishing planar frustration and cyclic exchange. 
\noindent It may be seen that LSW and SCSW theory show opposite tendency to each other. Though LSWT predicts the decrease in the magnetization with the grow of the parameter $\lambda_{\bot}$ due to the frustration which destabilizes the system, SCSW theory gives the growth in magnetization, as a consequence of the subtle interplay of the competing interactions. Besides, we notice that the same change in the parameters $\lambda_2$ and $\lambda_c$ yields within LSWT only a slight decrease of magnetization, while the increase within SCSWT is more significant.

\subsection{Other quantities}

Further, we calculate the spin-wave velocity renormalization factor by making use of expression (\ref{renorm_faktor}). We first increase the frustration ratio $\lambda_2$ and calculate the renormalization factor for different cyclic exchange ratios $\lambda_c$ in both LSWT and SCSWT approach (Figure 7a)). 

\bc{\includegraphics[scale=0.75]{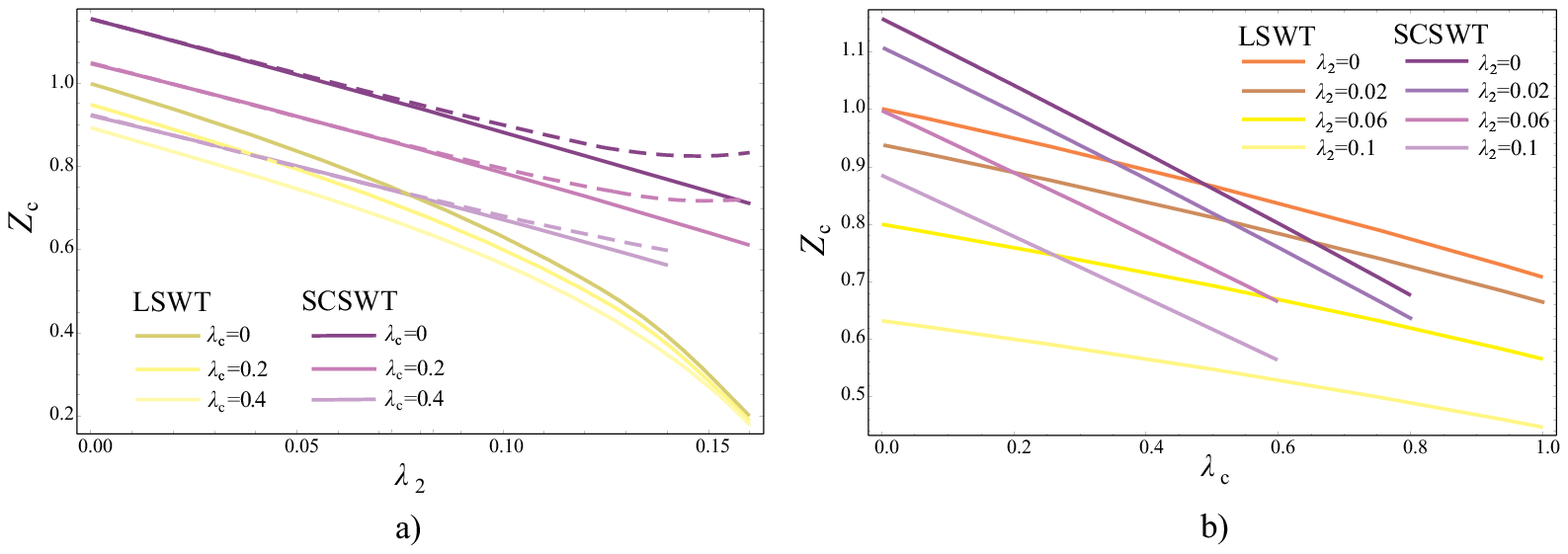} \\ {\small Figure 7: (Color online) a)Renormalization factor $Z_c$ for the spin-wave velocities vs. frustration $\lambda_2$ for different values of cyclic exchange $\lambda_c$, within LSWT and SCSWT approach. The dashed lines present the results of the first correction to LSWT. b)Renormalization factor $Z_c$ vs. $\lambda_c$, for different $\lambda_2$.}}\ec

\noindent The results demonstrate that $Z_c$ decreases with $\lambda_2$, the decrease being less rapid within the SCSW theory. The divergence similar to the one appearing in the analysis of $\langle \hat{S}^z\rangle_0$ is again eliminated within SCSWT. The renormalization factor is also calculated in dependence of $\lambda_c$ for different values of $\lambda_2$ (Figure 7b)), showing the more rapid decrease of $Z_c$ within SCSWT.

Finally, by making use of the expression (\ref{energ_osn_stanja}), we analyze the ground-state energy per lattice site versus frustration and cyclic exchange parameters. The dependence on the parameter $\lambda_2$ is plotted in Figure 8a). 

\bc{\includegraphics[scale=0.75]{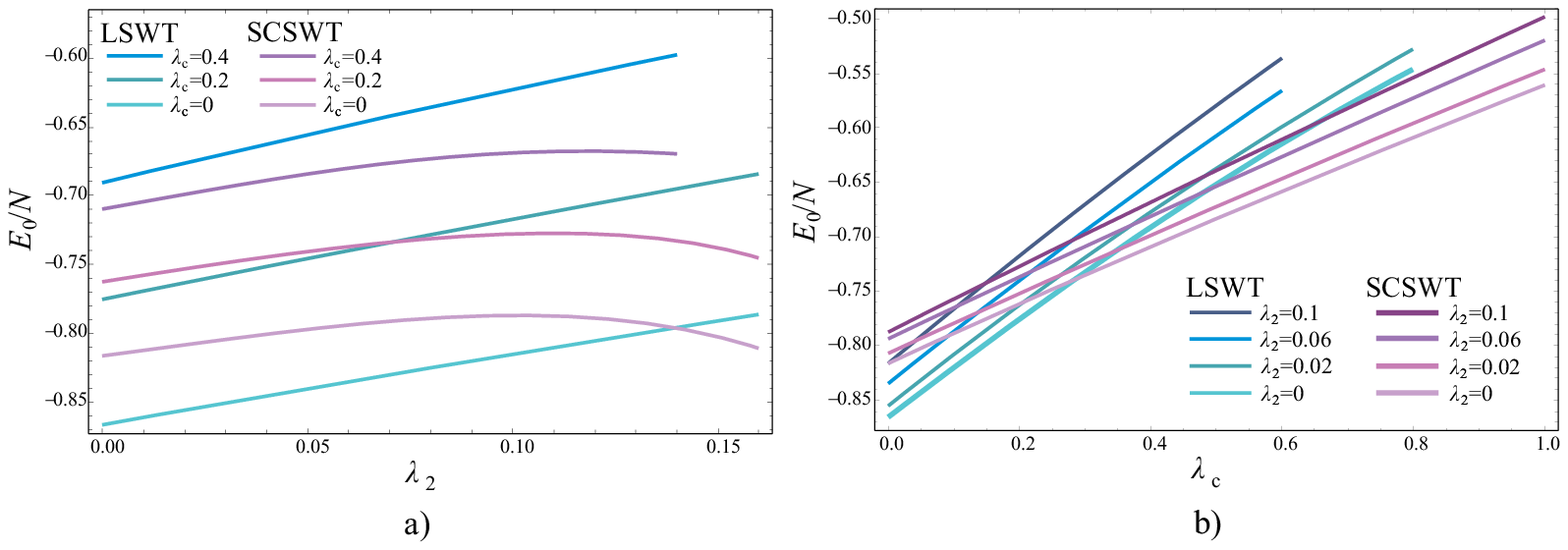} \\ {\small Figure 8:(Color online) a)Ground-state energy per lattice site vs. $\lambda_2$, for different $\lambda_c$, in LSW and SCSW approach. 
b)Ground-state energy per lattice site vs. $\lambda_c$, for different $\lambda_2$.}}\ec

\noindent It is interesting to notice that the absolute value of the ground-state energy within SCSWT at first decreases with frustration, however, after a certain value of the frustration ratio it starts to increase. Figure 8b) presents the dependence of the ground-state energy on the parameter $\lambda_c$. The results show that $|E_0/N|$ decreases with the growth of the cyclic exchange parameter. 

\subsection{Application to $\mbox{La}_2\mbox{CuO}_4$}

The model described by the Hamiltonian (\ref{hamiltonian1}) may be used to study the magnetic properties of the high-temperature superconductor parent compound $\mbox{La}_2\mbox{CuO}_4$. This compound has been the subject of our previous studies \cite{PRB1mi,PRB2mi,SSCmi}, where the calculation was performed within the method of spin Green functions on the model without the cyclic exchange interaction. The model with the non-vanishing parameter $\lambda_c$ within the SCSW theory has been examined in detail in Ref. \cite{starikatanin}. The set of bare superexchange couplings quoted therein was determined by the accurate fit of the in-plane spin-wave dispersion from Ref. \cite{staricoldea}, obtained by the inelastic neutron scattering at $10\,\mbox{K}$. Since we had improved experimental data from Ref. \cite{novicoldea} at our disposal, we recalculated the exchange parameters. The frustration parameter is taken to be as small as $\lambda_2=0.01$, according to \cite{majumdar_sa_cikl}, instead of the value $0.025$ from \cite{starikatanin}. The set of the exchange parameters is thus slightly changed and within the SCSW theory reads: $J=142.62\,\mbox{meV}$, $\lambda_c=0.22$ and $\lambda_2=\lambda_3=0.01$. If we use this set of parameters to obtain the spin-wave dispersion at $295\,\mbox{K}$, we obtain the spectrum which gives up to $25\%$ of the observed changes between the low- and high-temperature spectrum. This presents a mild improvement to the spectrum calculated from the parameters quoted in \cite{starikatanin}, since the later accounts only for a few percent of the observed changes. 

\section{Conclusions}

The three-dimensional tetragonal $S=1/2$ antiferromagnet presents an interesting system, studied in literature due to its application to layered copper oxides, which present the high-temperature superconducting parent compounds. In the present paper this system is described by the Heisenberg Hamiltonian comprising the planar interactions up to NNNN interaction, the cyclic exchange interaction, as well as the interplane interaction. The magnetic properties of the ground state of the model are examined within the framework of SCSW theory and compared to the LSWT predictions. 
\par We show that the system is highly frustrated and therefore subjected to the complex competition of different exchange interactions, leading to the specific magnetic properties behavior. The spin-wave spectra depend on frustration and cyclic exchange ratio, whereby the increase of these ratios yields the softening of the spectra. The sublattice magnetization dependence on the aforementioned parameters demonstrates that the zero-temperature quantum fluctuations play a pronounced role. The planar frustration and cyclic exchange have the opposite impact on the ground-state sublattice magnetization, whereby the former destabilizes, while the later stabilizes the system. The effect of the next-next-nearest neighbours interaction is shown to be significant, according to the fact that the critical value of the frustration ratio where AFM order is destroyed is, within LSWT for instance, three times smaller in the presence of the NNNN interaction. The influence of the interlayer coupling on the sublattice magnetization is also examined and turned out to be dictated by the exchange interaction competition. The spin-wave velocity renormalization factor and ground-state energy are strongly affected by the frustration and cyclic exchange interaction as well. 
\par The SCSW theory is also applied to layered cuprate oxide $\mbox{La}_2\mbox{CuO}_4$. The set of exchange parameters in this compound is determined to fit the experimental spin-wave dispersion obtained at $T=10\,\mbox{K}$ \cite{novicoldea}. The high-temperature spectrum is then calculated, showing better agreement with the experiment than the earlier ones, but still indicating to the limitations of the spin-wave theories at high temperatures.
\par The aim of this study was to give an exhaustive research of the results which SCSW theory gives for this model, as well as to point out its shortcomings, where some more sophisticated theories, as for example perturbative spin-wave expansion up to $1/S^2$, have to be applied.

%-----------------------------------------------------------------------------------
\subsection*{Acknowledgment} We are grateful to S. Hayden, R. Coldea and G. Aepply for sending us experimental data for the spin-wave  
dispersion in $\mbox{La}_2\mbox{CuO}_4$. We also thank Dr Slobodan Rado\v sevi\'c and Professor Milan Panti\'  c for helpful discussions. This work was supported by the Serbian Ministry of Education, Science and Technological Development, Project No. OI 171009.
%-----------------------------------------------------------------------------------


\begin{thebibliography}{10}

\bibitem{manusakis} E. Manousakis, Rev. Mod. Phys. {\bf 63}(1), (1991) 1-63. 
\bibitem{staricoldea} R. Coldea, S. M. Hayden, G. Aepply, T. G. Perring, C. D. Frost, T. E. Mason, S. W. Cheong, and Z. Fisk, Phys. Rev. Lett. {\bf 86} (2001) 5377. 
\bibitem{novicoldea} N. S. Headings, S. M. Hayden, R. Coldea, and T. G. Perring, Phys. Rev. Lett. {\bf 105} (2010) 247001. 
\bibitem{starikatanin} A. A. Katanin, and A. P. Kampf, Phys. Rev. B {\bf 66} (2002) 100403(R). 
\bibitem{majumdarIoPscsw} K. Majumdar, T. and Datta, J. Phys.: Condens. Matter {\bf 21} (2009) 406004.
\bibitem{majumdar_bez_cikl} K. Majumdar, Phys. Rev. B {\bf 82} (2010) 144407. 
\bibitem{majumdar_sa_cikl} K. Majumdar, D. Furton, and G. S. Uhrig G S, Phys. Rev. B {\bf 85} (2012) 144420.
\bibitem{majumdarIoPspinwaves} K. Majumdar, J. Phys.: Condens. Matter {\bf 23} (2011) 116004.
\bibitem{majumdarIoPmagnetization} K. Majumdar, J. Phys.: Condens. Matter {\bf 23}  (2011) 046001.
\bibitem{diep} H. T. Diep, \textit{Frustrated Spin Systems} 1st edn (Singapore: World Scientific) 2004.
\bibitem{zitomirski} M. E. Zhitomirsky, and A. L. Chernyshev, Rev. Mod. Phys. {\bf 85} (2013) 219.
\bibitem{stariirkin} V. Yu. Irkhin, A. A. Katanin, and M. I. Katsnelson, J. Phys.: Condens. Matter {\bf 4} (1992) 5227
\bibitem{irkin} V. Yu. Irkhin, A. A. Katanin, and M. I. Katsnelson, Phys. Rev. B {\bf 60} (1999) 1082.
\bibitem{kar} S. Kar, and T. Saha-Dasgupta, Physica B {\bf 432} (2014) 71-76. 
\bibitem{pavarini} E. Pavarini, S. C. Tarantino, T. B. Ballaran, M. Zema, P. Ghigna, and P. Carretta, Phys. Rev. B {\bf 77} (2008) 014425. 
\bibitem{irkinuspehi} A. A. Katanin, and V. Yu. Irkhin, Phys.-Usp. {\bf 50} (2007) 613.
\bibitem{PRB1mi} M. Manojlovi\' c, M. Pavkov, M. \v Skrinjar, M. Panti\' c, D. Kapor, and S. Stojanovi\' c, Phys. Rev. B {\bf 68} (2003) 014435. 
\bibitem{fetervaleka} A. L. Fetter, and J. D. Walecka, \textit{Quantum Theory of Many-Particle Systems} International Series in Pure and Applied Physics (New York: McGraw-Hill) 1971.
\bibitem{stanekPRB}D. Stanek, O. P. Sushkov, and G. S. Uhrig, Phys. Rev. B {\bf 84} (2011) 064505. 
\bibitem{flax}L. Flax, and J. C. Raich, Phys. Rev. B {\bf 2} (1970) 1339. 
\bibitem{PRB2mi}M. Rutonjski, S. Rado\v sevi\' c, M. \v Skrinjar, M. Pavkov-Hrvojevi\' c, D. Kapor, and M. Panti\' c, Phys. Rev. B {\bf 76} (2007) 172506. 
\bibitem{SSCmi} M. Rutonjski, S. Rado\v sevi\' c, M. Panti\' c, M. Pavkov-Hrvojevi\' c, D. Kapor, and M. \v Skrinjar, Solid State Commun. {\bf 151} (2011) 518-522. 


\end{thebibliography}
\end{document}